\documentclass[conference, a4paper]{IEEEtran}
\newcommand{\CLASSINPUTbottomtextmargin}{4.2cm}
\IEEEoverridecommandlockouts
\usepackage{cite}
\usepackage{amsmath,amssymb,amsfonts}
\usepackage{graphicx}
\usepackage{textcomp}
\usepackage{xcolor}
\usepackage{algorithm}
\usepackage{url}
\usepackage{xspace}
\usepackage{listings}
\usepackage[T1]{fontenc}
\usepackage{nicefrac}
\usepackage{siunitx}
\usepackage{array,framed}
\usepackage{booktabs}
\usepackage{
  color,
  float,
  epsfig,
  wrapfig,
  graphics,
  graphicx,
  subcaption
}

\usepackage{textcomp,amssymb}
\usepackage{setspace}
\usepackage{latexsym,fancyhdr,url}
\usepackage{enumerate}
\usepackage{algpseudocode}
\usepackage{graphics}
\usepackage{xparse} 
\usepackage{xspace}
\usepackage{multirow}
\usepackage{csvsimple}
\usepackage{balance}
\usepackage{
  tikz,
  pgfplots,
  pgfplotstable
}
\usepackage{mathtools,}
\usepackage{xspace}
\usepackage{caption}
\usepackage{subcaption}

\newcommand{\wenfei}[1]{{\color{red}{WW: #1}}}
\newcommand{\shaoke}[1]{{\color{blue}{SK: #1}}}
\newcommand{\qingsong}[1]{{\color{green}{QS: #1}}}
\newcommand{\sysname}{HyperNAT\xspace}

\def\BibTeX{{\rm B\kern-.05em{\sc i\kern-.025em b}\kern-.08em
    T\kern-.1667em\lower.7ex\hbox{E}\kern-.125emX}}
\begin{document}

\title{\sysname: Scaling Up Network Address Translation with SmartNICs for Clouds}

\author{\IEEEauthorblockN{Shaoke Fang\IEEEauthorrefmark{1}\textsuperscript{\textsection},
Qingsong Liu\IEEEauthorrefmark{2}\textsuperscript{\textsection}, and
Wenfei wu\IEEEauthorrefmark{3}\thanks{Wenfei Wu is the corresponding author.}}
\IEEEauthorblockA{
Tsinghua University\\
Email: \IEEEauthorrefmark{1}fsk19@mails.tsinghua.edu.cn,
\IEEEauthorrefmark{2}liu-qs19@mails.tsinghua.edu.cn,
\IEEEauthorrefmark{3}wenfeiwu@tsinghua.edu.cn}}

\maketitle
\begingroup\renewcommand\thefootnote{\textsection}
\footnotetext{Equal contribution}
\endgroup
\thispagestyle{plain}
\pagestyle{plain}

\begin{abstract}
Network address translation (NAT) is a basic functionality in cloud gateways. With the increasing traffic volume and number of flows introduced by the cloud tenants, the NAT gateway needs to be implemented on a cluster of servers. We propose to scale up the gateway servers, which could reduce the number of servers so as to reduce the capital expense and operation expense. We design \sysname, which leverages smartNICs to improve the server's processing capacity. In \sysname, the NAT functionality is distributed on multiple NICs, and the flow space is divided and assigned accordingly. \sysname overcomes the challenge that the packets in two directions of one connection need to be processed by the same NAT rule (named two-direction consistency, TDC) by cloning the rule to both data paths of the two directions. Our implementation and evaluation of \sysname show that \sysname could scale up cloud gateway effectively with low overhead.

\end{abstract}

\section{Introduction}

Modern public clouds support their tenants to build their own virtual private cluster (VPC) and connect with the Internet. A crucial component between the VPC and the external Internet is the \emph{cloud gateway}, which undertake the address translation between the VPC private addresses and the cloud public addresses. In current production networks, cloud gateways are usually implemented as software (e.g., Click, NetBrick, and DNAT\cite{kohler2000click,zhang2016opennetvm,panda2016netbricks,wu2011design}) on commodity servers for flexible deployment, routing, and management{\cite{qian2019flexgate}}.

The trend of migrating infrastructure into clouds for online server providers (e.g., {e-commerce, searching, e-retail, cloud computing, and multimedia entertainment}) leads to the exponential increase of cloud traffic, e.g., at an increasing rate of more than 60\% annually from 2015 to 2021~\cite{mlitz_2021}. However, the server computation power (CPU, PCIe, NIC) does not increase at the same speed. Until {2019}, a large data center needs to exchange 1Tbps traffic (peak 1.5Tbps) \cite{qian2019flexgate}, but the recent advanced Network Function Virtualization (NFV) platforms can only support processing traffic at a speed of 10-40Gbps \cite{gallo2018clicknf}\cite{jamshed2017mos}\cite{martins2014clickos}. As a result, cloud NAT gateway needs to be implemented in a distributed manner, usually on a rack of multiple servers.




Implementing one appliance (i.e., cloud gateway) on too many servers is not friendly to network management because that would lead to extra capital expense (CAPEX) and operation expense (OPEX) (e.g., complexity in availability and consistency control{\cite{qian2019flexgate}}). Thus, we propose to \emph{scale up the cloud gateway} instead of {scaling out} it --- {i.e., improving each cloud gateway server's processing capability.} We take the opportunity of recent progress in smartNICs; it has extra computation circuit on the NIC, providing the potential to offloading the NAT from the server to the NICs.

\sysname is designed based on the following observations. First, the processing capacity of a gateway server is not bounded by the NIC bandwidth(\textasciitilde
10Gbps  vs 25Gbps per NIC), but the datapath inside the server (NIC to memory, to CPU, and back to NIC; via bus). Second, the programmability of the smartNIC can handle the NAT logic, e.g., {Broadcom Stingray, Cisco Nexus X25\cite{cisconexussmartnic}, Mellanox Bluefield}\cite{mellanox}. Thus, we propose to offload one server's functionality/workload to its multiple smartNICs so that the bottleneck can be bypassed and the processing capacity can be increased proportionally to the number of NICs.

In \sysname, each gateway server is equipped with multiple smartNICs. The flow space of the server is partitioned into multiple subspaces, with each NIC in charge of one. Each flow is assigned to one NIC for the traffic processing by a hash function, i.e., address translation. Essentially, \sysname distributes the traffic to multiple NICs instead of the server CPU so that the processing capacity is scaled up.

We overcome the challenge of achieving two-direction consistency (TDC) for one flow among multiple NICs. For one connection, its outgoing traffic (from the tenant to the Internet) and incoming one has different flow identifiers (i.e., the five-tuple), and it is challenging to design a hash function that can assign two-direction packets to the same NIC. In \sysname, we do not try to assign the two-direction traffic to the same NIC; instead, we duplicate the state of a connection to both NICs (for its incoming and outgoing traffic), which achieves correctness in traffic processing.

We implement \sysname using Mellanox BlueField smartNIC and a commodity server. And evaluation shows that a prototype of \sysname with two smartNIC can scale up the processing capacity of one server by {1.28 to 1.60 times}. The latency and memory overhead are negligible.

\section{Background}
Cloud gateways are facing the scalability issue, and smartNICs provide the promise to scale up gateway servers.
\subsection{Cloud Gateway and Problem Statement}
\textbf{Functionality.} A cloud tenant usually sets up its virtual private cluster (VPC) inside the cloud and exchanges data with the Internet. The tenant can assign VMs in the VPC with virtual addresses (IP address and TCP port), whose space is called the \emph{internal address space} denoted by $I$. Different tenants' internal address spaces are isolated.

\begin{figure}
    \centering
    \includegraphics[width=\columnwidth]{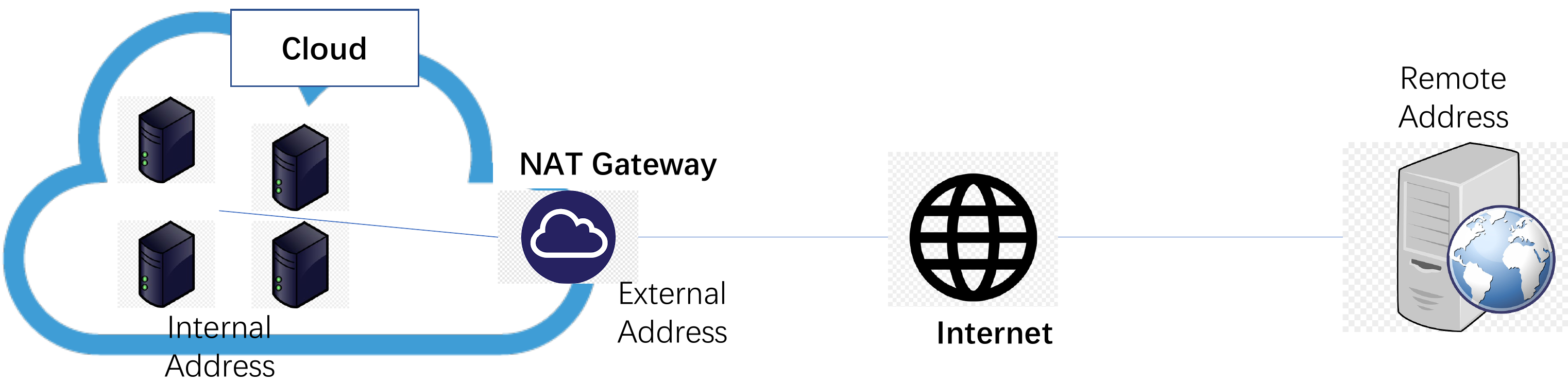}
    \caption{Three address spaces in traffic between VPC and the Internet}
    \label{fig:addr-space}
\end{figure}

When VMs in a cloud communicate with the Internet servers, their connection needs to be assigned with an external address. {The Internet routers use the external address and the Internet server address} to deliver traffic between the cloud and the Internet servers. The external space is denoted as $E$, and the remote Internet address is denoted as $R$. Figure~\ref{fig:addr-space} illustrates the three address spaces.

In the whole process, the internal VMs initiates a connection with source and destination addresses <$addr^{(I)}$, $addr^{(R)}$>, and the connection packets are translated by the cloud gateway to <$addr^{(E)}$, $addr^{(R)}$> and arrive at the Internet server; the Internet server replies traffic with addresses <$addr^{(R)}$, $addr^{(E)}$>, which is translated by the gateway to <$addr^{(R)}$, $addr^{(I)}$> and arrives at the internal VM. 

There are also connections initiated by cloud external servers to internal VMs, which is achieved by static assignment of a cloud external address (also called public address) to the internal VM address. This kind of address translation usually supports online services (a public address to clients), which has a separate component other than the one above supporting the internal-initiated connections. In this paper, we focus on the cloud gateway for internal-initiated connections.

NAT can be implemented as a piece of software or special hardware devices. Modern cloud usually choose software NAT for flexible management --- software NAT can be copied and booted up on any server, and they are easy to be duplicated as many instances to support the large volume of traffic.
 
\textbf{Scalability Issue.} The demand for cloud traffic increases much faster than the growth of processing capacity at the gateway server; thus, a cloud gateway is usually implemented on multiple servers in a cluster. By the report from Mlitz\cite{mlitz_2021}, the cloud traffic increases from {4.7 Zbps to 20.6 Zbps in five years.} But on the server side, CPU frequency does show an obvious increase in recent years, bus bandwidth takes several years to evolve one generation, and typical commodity NIC bandwidth increase from 10Gbps to 25Gbps.

As a result, the cloud provider needs to set up more instances of gateway to serve the traffic of increasing volume. Increasing servers is not in favor of network operators, as it causes obvious CAPEX. The OPEX is even worse, as the distributed implementation of one functionality on multiple servers usually introduces extra system complexity and lowers the availability (e.g., failover and consistency)\cite{huang2019proactive}.


\subsection{SmartNIC and Opportunity}
Recent progress in smartNICs inspires us to consider scaling up instead of scaling out per-server functionality. SmartNICs \cite{chen2005shangri,firestone2018azure,grant2020smartnic} usually have on-chip memory and can load and execute user-specified programs. The program could take the incoming packets and in-memory states as input, operate on both of them and send or drop packets. There are {three categories of smartNICs, which are based on ASIC, FPGA, and SoC technologies: ASIC-based has the fastest speed but with limited programmability\cite{chen2005shangri}; FPGA requires expert knowledge of the hardware to program but provides a performance near-ASIC\cite{firestone2018azure}; System-on-a-chip(SoC) has moderate computation power, but it is much easier to program\cite{grant2020smartnic}.} 

SmartNICs provide the promise to scale up a gateway server. According to our measurement, when a NAT server processes a stream of packets, the bottleneck is on the data path from arriving NIC to CPU and back to the NIC. Installing multiple smartNICs on the server and offloading the NAT logic can make the packet bypass the original data path and distributed it on multiple NICs. Even though each NIC may not be as powerful as the server, the multiple NICs together can exceed the server.

\subsection{Goal and Approach}
In this paper, our goal is to scale up a server with multiple smartNICs, guaranteeing the logical correctness as the on-server NAT and also achieving a throughput higher than that on the original server. Meanwhile, we also want the overhead to be negligible or acceptable.

We design a system named \sysname. \sysname duplicates the NAT logic on each smartNIC and distributed traffic among them. With per-flow consistent random hashing, each flow is assigned to one NIC, and all flows are expected to be assigned uniformly to multiple NICs. Then each NIC processes traffic within its flow space.

\textbf{Challenge.}
The challenge in implementing \sysname is the contradiction between keeping two-direction consistency (TDC) in NAT and the traffic load balancing mechanism. Each connection (e.g., TCP flow) contains packets in two directions --- we name the ones from the VPC to the Internet outgoing packets and the ones in the opposite direction incoming packets. The address translation rule of both directions much be consistent, i.e., outgoing packets having source address $addr^{(I)}$ translated to $addr^{(E)}$ and incoming packets having destination address $addr^{(E)}$ to $addr^{(I)}$. However, for a stateless load balancer using a hash function, it is difficult to hash <$addr^{(I)}$, $addr^{(R)}$> and <$addr^{(R)}$ $addr^{(E)}$> to the same smartNIC. 

Existing solutions (e.g., \cite{barbette2020high}) propose to attach a unique tag on the packets of a flow as the flow identifier and use the tag to direct the distribution of packets to processing units. This solution requires the modification of the end host networking stack so as to preserve the tag from the packet in one-direction to the return packet on the other, which is not always feasible. It also needs the router to support the routing number of flows (tags) at the cloud-scale, which is not practical either.

In \sysname, we abandon the design choice of keeping TDC at the same location. Instead, we allow the two-direction packets of one connection to be processed at different smartNICs. To keep the NAT rules consistent, when a rule has been installed for the first time, it will install on both smartNICs for packet in two directions. \S\ref{sec:Design} elaborate this design.

\section{Design}
\label{sec:Design}


\subsection{Overview}
\textbf{Functionality Distribution.} In \sysname, each smartNIC is installed with the NAT, but are in charge of the address translation in different flow space. Assume there are $N$ NICs, indexed from $1$ to $N$. The external address space $E$ is divided into $N$ non-overlapping subspaces, i.e., $E = E_1 \cup E_2 \cup ... \cup E_N$ and $E_i \cap E_j = \emptyset, \forall i\neq j$. The $i$-th NIC is in charge of the external address space $E_i$.

\textbf{Workflow.} The server is at the edge between the cloud and the Internet. Each of the server's NIC is connected to a port of the server's access switch. For one outgoing packet (from switch to the server), the switch would hash (e.g., ECMP) its address <$addr^{(I)}$, $addr^{(R)}$> to one of the link/NIC. Assume the $i$-th NIC is chosen by the hash function to process the packet, and then the address is translated to <$addr^{(E_i)}$, $addr^{(R)}$>. The NATed packet is sent to the Internet server ($addr^{(R)}$). 

The return incoming packet of the same connection would have the address <$addr^{(R)}$, $addr^{(E_i)}$>. The switch would hash the return packet to one of the NICs. Assume the $j$-th NIC is chosen for the return packet. Then it needs to translate the address <$addr^{(R)}$, $addr^{(E_i)}$> back to <$addr^{(R)}$, $addr^{(I)}$>.

As discussed in the previous section, the hash function cannot guarantee $j=i$, so the incoming packets may need to be processed by the $j$-th NIC which does not contain the address translation rule ($addr^{(E_i)}$ to $addr^{(I)}$). Existing work tries to install routing rules in the switch to make the incoming packet routed to the $i$-th NIC again, which is not scalable for the cloud-scale ($10^5$ active flows).

Instead, \sysname would clone the address translation rule from the $i$-th NIC to the $j$-th NIC when the first packet triggers NAT. For the following packets, the outgoing ones would still be processed by the $i$-th NIC, and the incoming ones would be processed by the $j$-th, which is already installed with the address translation rule.

There are two options for the rule cloning --- active installation and passive installation. In the passive installation, when the $j$-th NIC meets with an unseen packet, it uses the source address $addr^{(E_i)}$ to find that the rule is in the $i$-th NIC and fetch the rule back. In the active installation, when the $i$-th NIC first creates the rule, it uses the same hash function as in the switch to compute the NIC of the return packets (on $j$-th NIC) and send the rule to the $j$-th NIC.

\sysname chooses passive installation due to resource limitations. Passive rule installation would install rules on demand, which could save or amortize resources, including the bandwidth between two NICs and the memory/rule entries on the target NIC. In our experiment, we observe that the bandwidth between two NICs is the system bottleneck. Thus, we choose the passive rule installation.




 
 
 \begin{figure}[htp]
    \centering
    \includegraphics[width=\columnwidth]{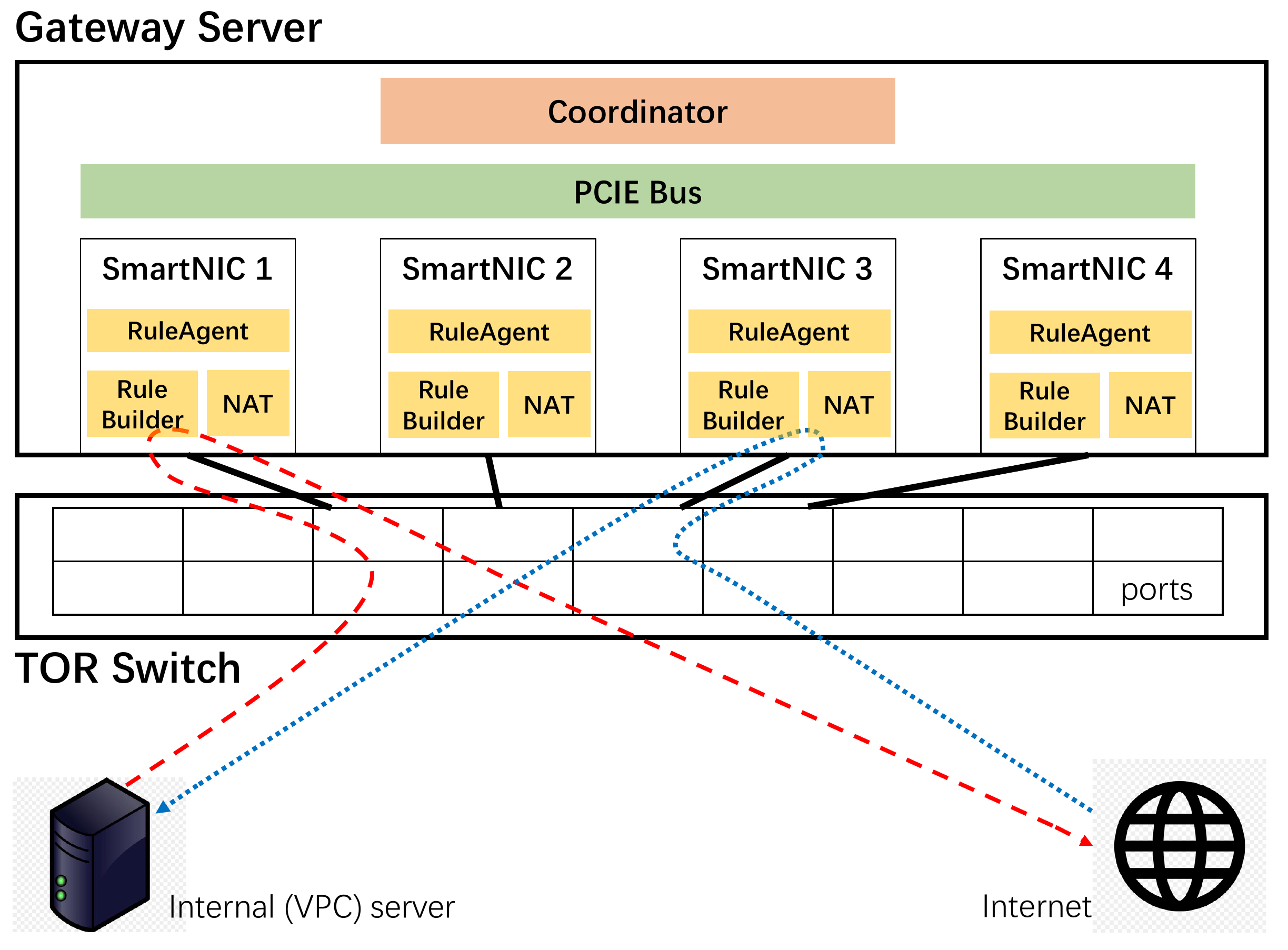}
    \caption{\sysname architecture and an example connection (two-direction traffic)}
    \label{fig:Design-Architecture}
\end{figure}

\subsection{Logic in Each Module}
Figure~\ref{fig:Design-Architecture} shows the system architecture. Each smartNIC has three modules --- NAT, RuleBuilder, and RuleAgent to perform its logic. Furthermore, the server has a coordinator for rule exchange between NICs. The server, together with the smartNICs, exchange traffic with the access switch and perform NAT functionality.

\textbf{RuleBuilder.} The RuleBuilder is in charge of creating address translation rules, and it maintains the external address space $E_i$. When the first (outgoing) packet of a connection initiates the connection, it is not found in the NAT, and the RuleBuilder would use find an empty address in $E_i$ and assign it to the flow. The rule is installed on NAT, and the packet is sent back to NAT for further processing. 

Meanwhile, \sysname adopts active rule installation on the return path, so the rule is sent to RuleAgent for rule exchange. The message format is {\tt <target\_NIC\_ID, rule>}.

\textbf{NAT.} NAT would not take charge of rule creation but only perform address translation according to the rules assigned to it from the RuleBuilder. When a packet does not match any rules, indicating it is the first packet of a connection, the NAT module will send it to the RuleBuilder.

NAT would accept rule installation from the RuleBuilder and perform address translation for all packets which match a rule in it.

\textbf{RuleAgent.} RuleAgent connects with the coordinator on the server. When receiving a rule message from the RuleBuilder, it would forward the message to the coordinator. When receiving a rule message from the RuleBuilder, it would install the rule into the local NAT.

\textbf{Coordinator.} The coordinator work as a router for NAT rule to exchange rules between NICs. When receiving a message of {\tt <target\_NIC\_ID, rule>}, it would parse and forward the message to the target NIC.



\subsection{Analysis of Availability}
Partitioning the external space to subspaces on NICs could reduce the system availability, and we regard it as a tradeoff for the system throughput. Using the notation above, assume there are $X$ flows to be assign to an address space of size $F$ ($F=|E|$). The $X$ flows are randomly assigned to the $N$ NICs. For each NIC, the number of flows that is assigned to it is a variable $x$, and its expectation $E(x)$ is $X/N$. Using Markov inequality, we have 
\begin{equation}
    Pr[x > F/N] \leq E(x)/(F/N).
\end{equation}
Thus, the probability that one NIC's address space is overflown is 
\begin{equation}
    Pr[x > F/N] \leq (X/N)/(F/N) = X/F.
\end{equation}
The probability that one of the $N$ NICs experiencing overflow is 
\begin{equation}
     1-(1-X/F)^N\approx XN/F.
\end{equation}

With more NICs ($N$), the system is more like to fail to serve a flow. Nevertheless, in practice, when serving 10K simultaneous active users (each with 10 simultaneous active flows), $X$ is estimated to be $10^5$, $F$ is larger than $10^9$ (a B class subnet, each IP with 65536 ports), and $N$ is between 1 and 10. The failure rate is less than 0.1\%, which is negligible.

\section{Evaluation}

\subsection{Implementation and Experiment Settings}

\textbf{Implementation}
We implement \sysname on Mellanox {Bluefield (1st generation)} NICs, which has ARM-based network processors. \sysname is implemented in C/C++, where the three modules on the smartNIC have 3800 lines of code, and the coordinator takes 600 lines.

\begin{figure}[htp]
    \centering
    \begin{subfigure}[b]{0.49\columnwidth}
         \centering
         \includegraphics[width=\textwidth]{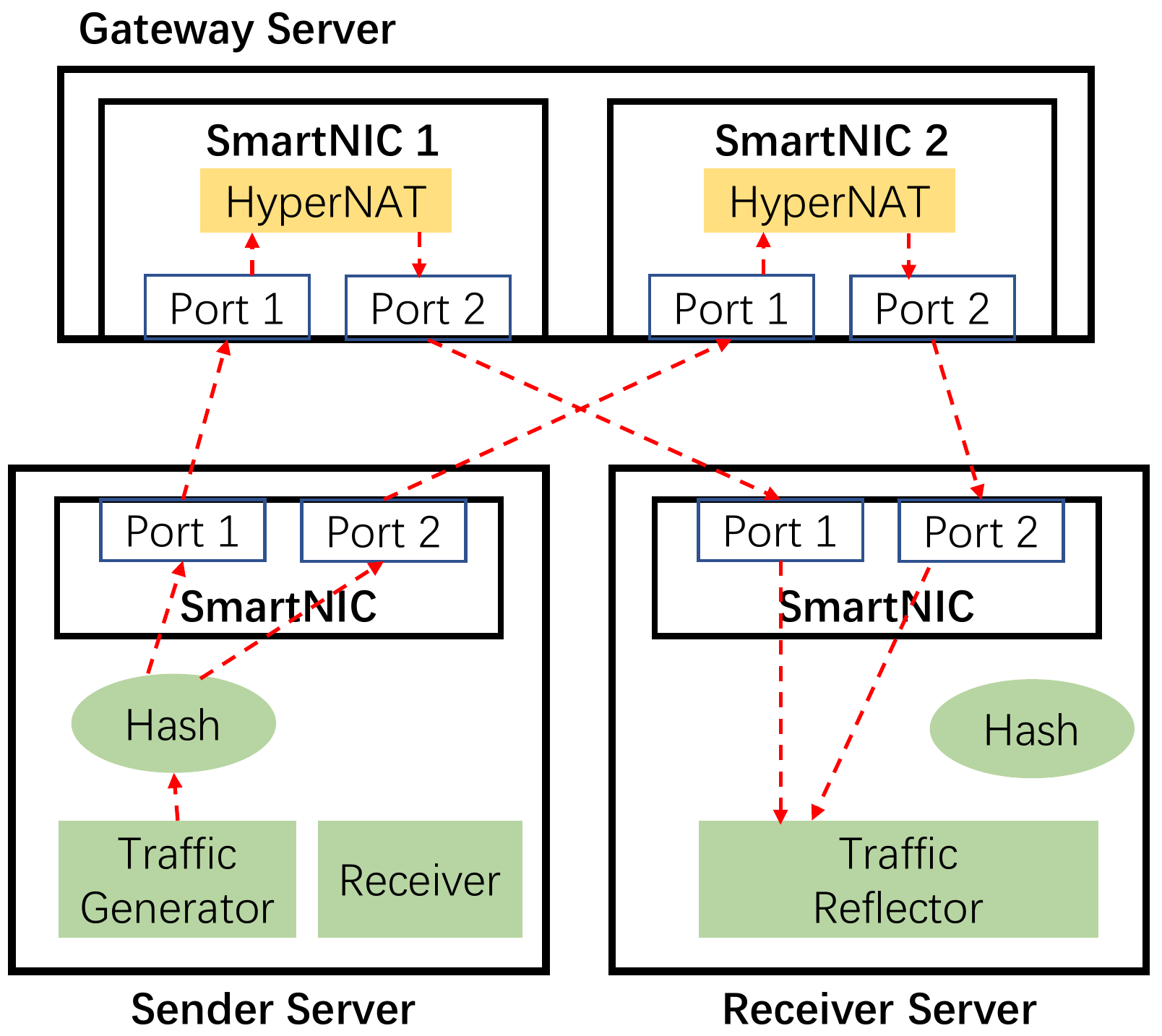}
         \caption{Emulate outgoing flows}
         \label{fig:exp-outgoing}
     \end{subfigure}
     \hfill
     \begin{subfigure}[b]{0.49\columnwidth}
         \centering
         \includegraphics[width=\textwidth]{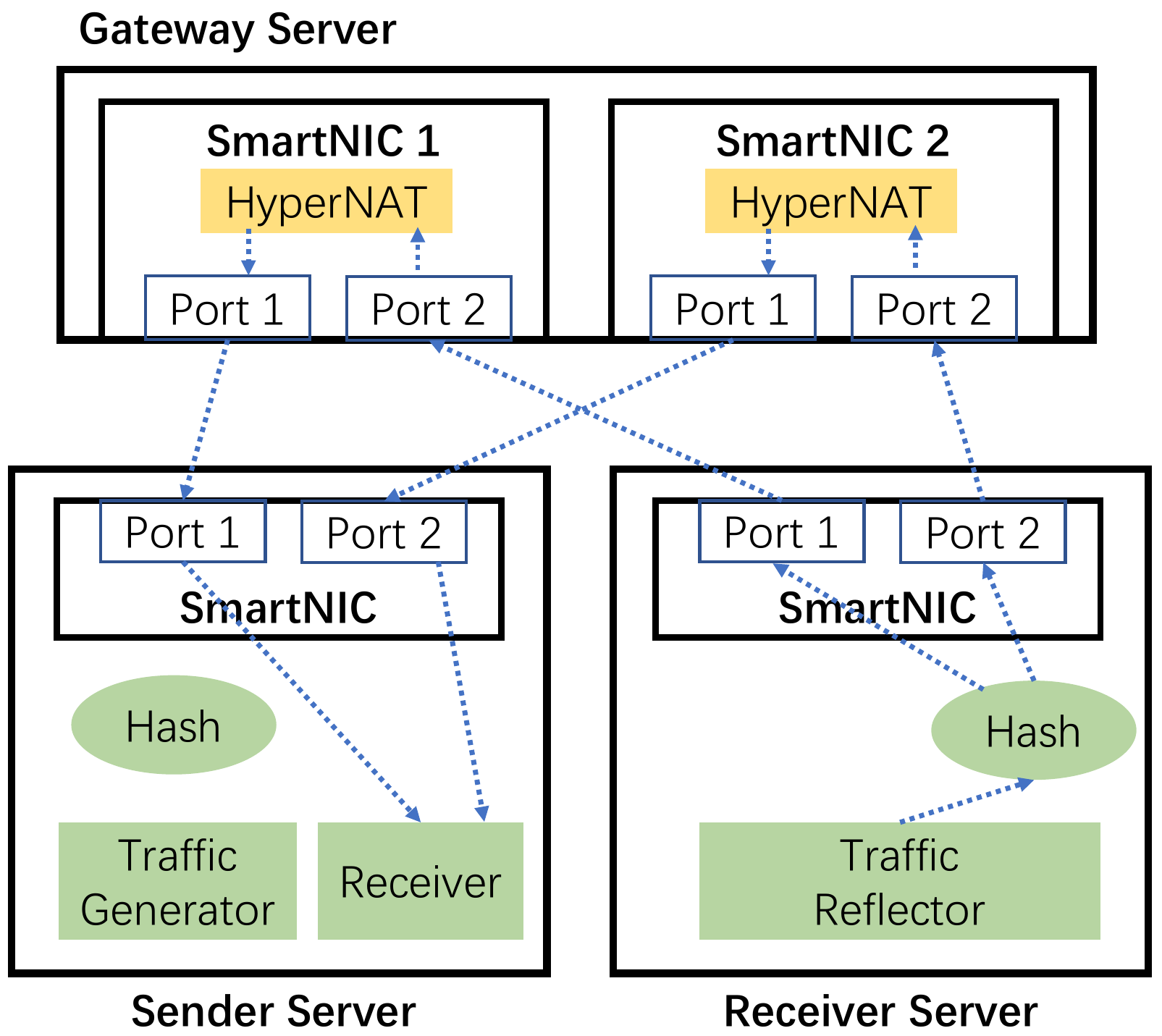}
         \caption{Emulate incoming flows}
         \label{fig:exp-incoming}
     \end{subfigure}
    \caption{Topology in experiments}
    \label{fig:Evaluation-Architecture}
\end{figure}

\textbf{Environment and Topology.} In our experiment, we use workstations of {8}-core {3.6} GHz CPU and {16}GB memory. All the NICs are Mellanox {25}Gbps dual port NICs.

The experiment topology is shown in Figure~\ref{fig:Evaluation-Architecture}. We use one server that emulates the access switch, which sends traffic. The sender server's dual-port NIC has two cables connected with two NICs on the NAT server, and it generates traffic and applies a SHA-1 hash function to load balance traffic to two ports.

The NAT server is equipped with two NICs, with each NIC running the three NIC NAT modules and the server running the coordinator. Each has one port connecting with the sender NIC and another connecting with the receiver server.

The receiver server emulates the Internet server, which has a dual-port NIC. The receiver would fetch incoming packets from both ports, reverse their source and destination address, use the same hash function as the sender side to choose one of the two ports, and send the packets back.

\textbf{Packet Trace.} We use the trace collected from a university in our experiment\cite{benson2010network} . From the trace, we randomly pick 10k, 50k, 100k, and 200k flows (extracting their packets) and use them as Trace 1, 2, 3, and 4 respectively. In our experiment, we make the sender deliver traffic at 2.0 MPPS to saturate the data path and measure the metrics of ``processed'' traffic. 

\textbf{Baselines.} We compare three architectures. The default \sysname has two smartNICs, and \sysname(1NIC) has only one NIC enabled. The serverNAT is the existing solution, where the server runs NAT and uses only one NIC for network I/O.

\textbf{Metrics.} We measure the throughput to show the system processing capacity. We measure the resource utilization and packet latency to show its overhead. Moreover, we also simulate the overloaded scenario to evaluate the system availability.


\subsection{Performance and Overhead}
We validate the NAT functionality by checking (1) different flows are assigned with different addresses, and (2) packets of one connection on two directions follow the same address translation (between the internal address space and the external one). \sysname passes all these tests with 100\% correctness. We further profile its performance.

\begin{figure}[htp]
    \centering
    \includegraphics[width=\columnwidth]{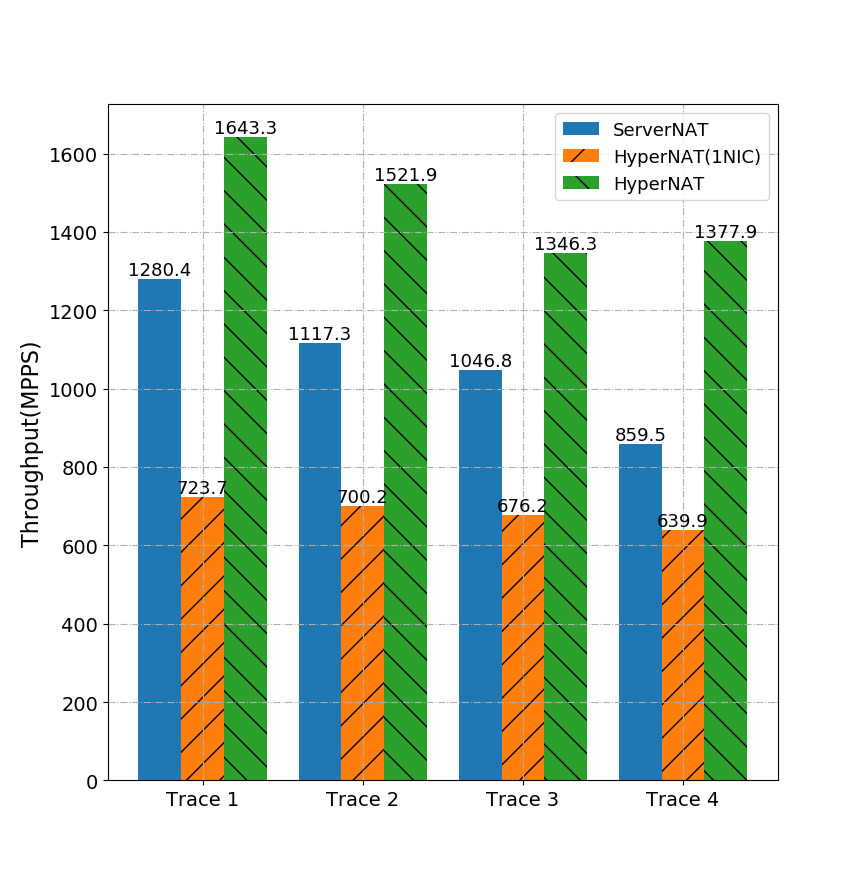}
    \caption{Throughput of \sysname and two baselines, varying traces}
    \label{fig:Throughput}
\end{figure}

\textbf{Throughput.} Figure~\ref{fig:Throughput} shows the system throughput comparing \sysname with two baselines and varying the traces. We get the following observations. First, \sysname outperforms serverNAT, which further outperforms \sysname(1NIC). For example, with trace 1, \sysname, serverNAT, and \sysname(1NIC) achieve a throughput of {1643.3Kpps, 1280.4Kpps, and 723.7Kpps} respectively. The reason is that the network processor has less computation power than the server CPU, but \sysname distributes the computation on two NICs, who together exceed the power of a single server CPU. Second, with the number of flows increasing in the trace, all three solutions' throughput decrease (e.g., from 1643.3Kpps to 1377.9Kpps for \sysname, and 1280.4Kpps to 859.5Kpps for serverNAT). This happened because each flow's first packet needs a rule creation, whose total cost is proportional to the number of flows. In \sysname, this decreasing trend is more significant; because the \sysname's return packets have a passive rule installation process, whose data flow crosses two NICs and the server.

\begin{figure}[htp]
    \centering
    \includegraphics[width=\columnwidth]{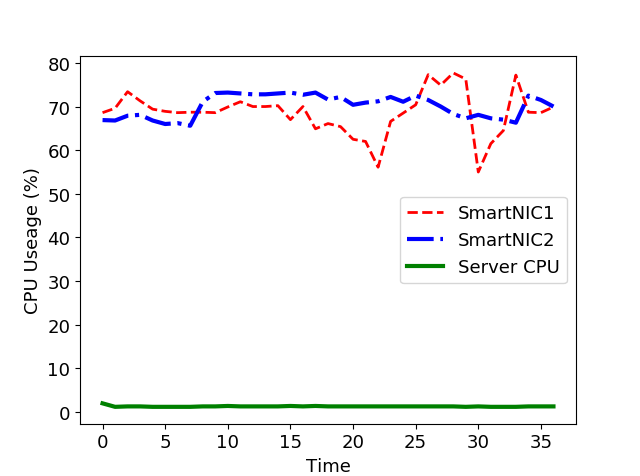}
    \caption{CPU usage with trace 1}
    \label{fig:Cpu-usage}
\end{figure}

\textbf{CPU Usage.} Figure~\ref{fig:Cpu-usage} shows the CPU usage of two smartNICs and the server when running the \sysname experiment with Trace 1. We observe that the NAT computation workload is offloaded to the two NICs (each about 70\%), and the server CPU is idle (less than 2\%). Thus, \sysname could save CPU resources for servers.


\textbf{Latency.} \sysname performs passive rule installation, which may cause the first return packet to suffer from rule fetching round-trip waiting. We measure this latency overhead.

Table~\ref{tab:Event-timestamp} shows the timestamps of events for a packet and its ACK. The experiment is repeated 200 times, and the average timestamp and the standard deviation are computed in the table. We observe that the first return packet would cause about 2.3ms latency. Once the rule is installed, this latency is eliminated for the following packets. This latency is acceptable compared with the round-trip time (RTT) on the Internet, which is about tens of milliseconds (the first packet are even longer on the Internet most of the time).

\begin{table}[htb]
\centering
\caption{Event timestamps when packets traverse \sysname}
\label{tab:Event-timestamp}
\small
\begin{tabular}{|c|c|c|}
\hline
\textbf{Event}                & \textbf{1st pkt/ACK} & \textbf{other pkts/ACK} \\ \hline
Start at sender           & 0                    & 0                          \\ \hline
Arrive at 1st NIC         & 100us   & 100us                      \\ \hline
Rule installed on 1st NIC & 125us                & 100us                      \\ \hline
Receiver relieved packet      & 225us                & 207us                        \\ \hline
Receiver sent out ACK         & 325us                & 311us                        \\ \hline
ACK at 2nd NIC            & 427us                & 411us                     \\ \hline
Rule installed on 2nd NIC & 2168us               & 411us                        \\ \hline
Return to sender              & 2269us              & 621us                        \\ \hline
\end{tabular}
\end{table}

\begin{figure}[htp]
    \centering
    \includegraphics[width=\columnwidth]{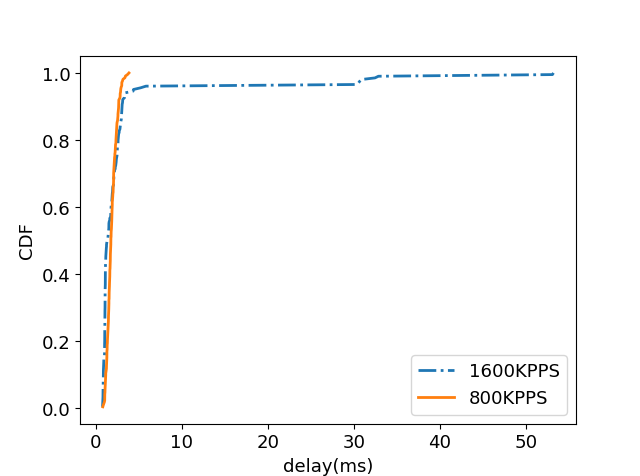}
    \caption{CDF of RTT, varying traffic rate}
    \label{fig:RTT-CDF}
\end{figure}

Figure~\ref{fig:RTT-CDF} shows the CDF of the RTT for packets in \sysname experiments (trace 1, varying sending rate to be 800KPPS and 1600KPPS). When the workload is light, the RTT is low (98.6\%-ile is 3.0ms for 800KPPS). When the workload is heavy, the tail latency is more significant. 
3.4\% of packets would experience RTT longer than 30.3ms. Most first packets would suffer more than in the scenario of light workload; a few non-frst packets would also be affected because the heavy workload makes packet arrive in larger density (and being affected more).

\section{Related Work}

\textbf{NAT.} NAT can be implemented in hardware or software~\cite{kohler2000click,zhang2016opennetvm,panda2016netbricks}. Cloud gateway choose software implementation for flexible management. DNAT is a design of distributed NAT\cite{wu2011design} using centralized address management (application and allocation), which may have extra overhead for address (de)allocation. HyperNAT partition the address space so that each smartNIC runs independently.

\textbf{SmartNIC Offloading.} A class of literature present the method and management framework to offload packet processing functionality onto smartNICs, e.g., UNO~\cite{le2017uno} and UDT~\cite{sabin2015security}. Durner et. al. proposes to distribute traffic between NICs and the host according to flow characteristics~\cite{durner2020network}. FlexGate~\cite{qian2019flexgate} proposes to offload proper functionalities among diverse ones to the NIC and the remaining are on the end host. FairNIC~\cite{grant2020smartnic} is a solution to isolate tenants' traffic on the NIC. Compared with these works, \sysname offload simple functionalities and randomly distribute traffic among NICs; its packet processing logic is not complex, but it extraly handles the per-connection consistency problem.

\textbf{Connection Consistency.} SilkRoad~\cite{miao2017silkroad} and Cheetah~\cite{barbette2020high} solve the per-connection consistency problem, i.e., when the address space among multiple processing units (servers in these two works and smartNICs in \sysname) is dynamically adjusted, existing flows should preserve its existing processing location. In \sysname, the load balancing design propose a new consistency requirement --- two direction consistency (TDC). Packets on two directions of the same connection should be processed by the same rule. And \sysname achieves this by duplicating the rules among smartNICs.

\textbf{State/Rule Migration.} Like several existing works that needs migrate flow/states among multiple running instances, e.g., OpenNF~\cite{gember2014opennf}, and they have complicated mechanisms to guarantee properties of loss freedom and order preservation. In HyperNAT, the rule clone is in the beginning of a flow, saving the troubles of keeping these properties.



\section{Conclusion}
We designed \sysname, which scale up the cloud NAT gateway server. The server is equipped with multiple smartNICs, with each of which processing a non-overlapping set of flows. \sysname overcomes the challenge of two-direction consistency by cloning NAT rules on both paths of a connection's two directions. Our prototype and evaluation shows that \sysname outperforms a single server in terms of the processing capacity and latency.

\bibliographystyle{plain}
\bibliography{refs}

\begin{thebibliography}{10}

\bibitem{mellanox}
Bluefield™ smartnic ethernet.
\newblock \url{https://www.mellanox.com/products/BlueField-SmartNIC-Ethernet}.

\bibitem{cisconexussmartnic}
cisco nexus x25 smartnic k3p-s data sheet\_2021.
\newblock
  \url{https://www.cisco.com/c/en/us/products/collateral/interfaces-modules/nexus-smartnic/datasheet-c78-743827.html},
  Jan 2021.

\bibitem{barbette2020high}
Tom Barbette, Chen Tang, Haoran Yao, Dejan Kosti{\'c}, Gerald~Q Maguire~Jr,
  Panagiotis Papadimitratos, and Marco Chiesa.
\newblock A high-speed load-balancer design with guaranteed
  per-connection-consistency.
\newblock In {\em 17th $\{$USENIX$\}$ Symposium on Networked Systems Design and
  Implementation ($\{$NSDI$\}$ 20)}, pages 667--683, 2020.

\bibitem{benson2010network}
Theophilus Benson, Aditya Akella, and David~A Maltz.
\newblock Network traffic characteristics of data centers in the wild.
\newblock In {\em Proceedings of the 10th ACM SIGCOMM conference on Internet
  measurement}, pages 267--280, 2010.

\bibitem{chen2005shangri}
Michael~K Chen, Xiao~Feng Li, Ruiqi Lian, Jason~H Lin, Lixia Liu, Tao Liu, and
  Roy Ju.
\newblock Shangri-la: Achieving high performance from compiled network
  applications while enabling ease of programming.
\newblock {\em ACM SIGPLAN Notices}, 40(6):224--236, 2005.

\bibitem{durner2020network}
Raphael Durner and Wolfgang Kellerer.
\newblock Network function offloading through classification of elephant flows.
\newblock {\em IEEE Transactions on Network and Service Management},
  17(2):807--820, 2020.

\bibitem{firestone2018azure}
Daniel Firestone, Andrew Putnam, Sambhrama Mundkur, Derek Chiou, Alireza
  Dabagh, Mike Andrewartha, Hari Angepat, Vivek Bhanu, Adrian Caulfield, Eric
  Chung, et~al.
\newblock Azure accelerated networking: Smartnics in the public cloud.
\newblock In {\em 15th $\{$USENIX$\}$ Symposium on Networked Systems Design and
  Implementation ($\{$NSDI$\}$ 18)}, pages 51--66, 2018.

\bibitem{gallo2018clicknf}
Massimo Gallo and Rafael Laufer.
\newblock Clicknf: a modular stack for custom network functions.
\newblock In {\em 2018 $\{$USENIX$\}$ Annual Technical Conference
  ($\{$USENIX$\}$$\{$ATC$\}$ 18)}, pages 745--757, 2018.

\bibitem{gember2014opennf}
Aaron Gember-Jacobson, Raajay Viswanathan, Chaithan Prakash, Robert Grandl,
  Junaid Khalid, Sourav Das, and Aditya Akella.
\newblock Opennf: Enabling innovation in network function control.
\newblock {\em ACM SIGCOMM Computer Communication Review}, 44(4):163--174,
  2014.

\bibitem{grant2020smartnic}
Stewart Grant, Anil Yelam, Maxwell Bland, and Alex~C Snoeren.
\newblock Smartnic performance isolation with fairnic: Programmable networking
  for the cloud.
\newblock In {\em Proceedings of the Annual conference of the ACM Special
  Interest Group on Data Communication on the applications, technologies,
  architectures, and protocols for computer communication}, pages 681--693,
  2020.

\bibitem{huang2019proactive}
Huawei Huang and Song Guo.
\newblock Proactive failure recovery for nfv in distributed edge computing.
\newblock {\em IEEE Communications Magazine}, 57(5):131--137, 2019.

\bibitem{jamshed2017mos}
Muhammad~Asim Jamshed, YoungGyoun Moon, Donghwi Kim, Dongsu Han, and KyoungSoo
  Park.
\newblock mos: A reusable networking stack for flow monitoring middleboxes.
\newblock In {\em 14th $\{$USENIX$\}$ Symposium on Networked Systems Design and
  Implementation ($\{$NSDI$\}$ 17)}, pages 113--129, 2017.

\bibitem{kohler2000click}
Eddie Kohler, Robert Morris, Benjie Chen, John Jannotti, and M~Frans Kaashoek.
\newblock The click modular router.
\newblock {\em ACM Transactions on Computer Systems (TOCS)}, 18(3):263--297,
  2000.

\bibitem{le2017uno}
Yanfang Le, Hyunseok Chang, Sarit Mukherjee, Limin Wang, Aditya Akella,
  Michael~M Swift, and TV~Lakshman.
\newblock Uno: Uniflying host and smart nic offload for flexible packet
  processing.
\newblock In {\em Proceedings of the 2017 Symposium on Cloud Computing}, pages
  506--519, 2017.

\bibitem{martins2014clickos}
Joao Martins, Mohamed Ahmed, Costin Raiciu, Vladimir Olteanu, Michio Honda,
  Roberto Bifulco, and Felipe Huici.
\newblock Clickos and the art of network function virtualization.
\newblock In {\em 11th $\{$USENIX$\}$ Symposium on Networked Systems Design and
  Implementation ($\{$NSDI$\}$ 14)}, pages 459--473, 2014.

\bibitem{miao2017silkroad}
Rui Miao, Hongyi Zeng, Changhoon Kim, Jeongkeun Lee, and Minlan Yu.
\newblock Silkroad: Making stateful layer-4 load balancing fast and cheap using
  switching asics.
\newblock In {\em Proceedings of the Conference of the ACM Special Interest
  Group on Data Communication}, pages 15--28, 2017.

\bibitem{mlitz_2021}
Kimberly Mlitz.
\newblock Global data center ip traffic 2013-2021, Jan 2021.

\bibitem{panda2016netbricks}
Aurojit Panda, Sangjin Han, Keon Jang, Melvin Walls, Sylvia Ratnasamy, and
  Scott Shenker.
\newblock Netbricks: Taking the v out of $\{$NFV$\}$.
\newblock In {\em 12th $\{$USENIX$\}$ Symposium on Operating Systems Design and
  Implementation ($\{$OSDI$\}$ 16)}, pages 203--216, 2016.

\bibitem{qian2019flexgate}
Kun Qian, Sai Ma, Mao Miao, Jianyuan Lu, Tong Zhang, Peilong Wang, Chenghao
  Sun, and Fengyuan Ren.
\newblock Flexgate: High-performance heterogeneous gateway in data centers.
\newblock In {\em Proceedings of the 3rd Asia-Pacific Workshop on Networking
  2019}, pages 36--42, 2019.

\bibitem{sabin2015security}
Gerald Sabin and Mohammad Rashti.
\newblock Security offload using the smartnic, a programmable 10 gbps ethernet
  nic.
\newblock In {\em 2015 National Aerospace and Electronics Conference (NAECON)},
  pages 273--276. IEEE, 2015.

\bibitem{wu2011design}
Zhigang Wu, Hao Luo, Shuzhuang Zhang, and Tao Zhang.
\newblock Design of a distributed network address translation system
  architecture.
\newblock 2011.

\bibitem{zhang2016opennetvm}
Wei Zhang, Guyue Liu, Wenhui Zhang, Neel Shah, Phillip Lopreiato, Gregoire
  Todeschi, KK~Ramakrishnan, and Timothy Wood.
\newblock Opennetvm: A platform for high performance network service chains.
\newblock In {\em Proceedings of the 2016 workshop on Hot topics in Middleboxes
  and Network Function Virtualization}, pages 26--31, 2016.

\end{thebibliography}

\end{document}